\documentclass[a4paper,fleqn,usenatbib]{mnras}

\usepackage[T1]{fontenc}
\usepackage{ae,aecompl}

\usepackage{graphicx}	
\usepackage{amsmath}	
\usepackage{amssymb}	
\usepackage{pdflscape} 
\usepackage{afterpage}
\usepackage{amsmath}
\usepackage{bm}

\title{Identifying MgII Narrow Absorption Lines with Deep Learning}

\author[Y. Zhao et al.]{
Yinan Zhao,$^{1}$\thanks{E-mail: yinanzhao@ufl.edu}
Jian Ge,$^{1}$
Xiaoyong Yuan,$^{2}$
Tiffany Zhao,$^{1}$
Cindy Wang,$^{1}$
Xiaolin Li$^{2}$
\\
$^{1}$Department of Astronomy, University of Florida, Gainesville, FL, 32611\\
$^{2}$National Science Foundation Center for Big Learning, University of Florida, Gainesville, FL, 32611\\
}

\date{Accepted XXX. Received YYY; in original form ZZZ}

\pubyear{2017}

\begin{document}
\label{firstpage}
\pagerange{\pageref{firstpage}--\pageref{lastpage}}
\maketitle

\begin{abstract}
Metal absorption line systems in distant quasar spectra probe of the history of gas content in the 
universe. The MgII $\lambda \lambda$ 2796, 2803 doublet is one of the most important absorption lines since it is 
a proxy of the star formation rate and a tracer of the cold gas associated with high redshift galaxies.
 Machine learning algorithms have been used to detect absorption lines systems
 in large sky surveys, such as Principle Component Analysis (PCA), Gaussian Process (GP) and decision trees. 
A very powerful algorithm in the field of machine learning called deep neural networks, or `` deep learning'' 
is a new structure of neural network that automatically extracts semantic features from raw data and
 represents them at a high level. In this paper, we apply a deep convolutional neural network for absorption 
line detection. We use the previously published DR7 MgII catalog  (Zhu et al. 2013) as the training and validation sample 
and the DR12 MgII catalog as the test set. Our deep learning algorithm is capable of detecting MgII absorption 
lines with an accuracy of $\sim$94\% . It takes only  $\sim 9$ seconds to analyze $\sim$ 50000 quasar spectra with our 
deep neural network, which is ten thousand times faster than traditional methods, while preserving high accuracy with
 little human interference. Our study shows that Mg II absorption line detection accuracy of a deep neutral network model strongly depends on the filter size in the filter layer of the neural network, and the best results are obtained when the filter size closely matches the absorption feature size.
 \end{abstract}


\section{Introduction}

Intervening quasar absorption line systems are a powerful tool to investigate gas kinematics and galaxy evolution from
 the early universe to present day.  The MgII absorption doublet
 is particularly useful, since its wavelengths  $\lambda \lambda$ 2796, 2803 can be detected in quasar optical spectra
 to trace galaxies in the intermediate redshift range ($0.3 < z < 2.5$), 
which covers the peak of global star formation history
in the universe (Madau \& Dickinson 2014 and references therein).
 The Sloan Digital Sky Survey (SDSS; York et al. 2000) contains 
a large sample of more than 400,000 quasar spectra well-suited for the investigation of MgII absorber systems. 
Mg II absorption doublets have been detected in early SDSS quasar spectral data releases with various completeness
 using traditional detection methods. For instance,  Nestor et al.(2005) identified 
over 1300 doublets in the SDSS Early Data Release. The completeness of their sample 
is $\sim 97\%$ based on their Monte Carlo simulation. Zhu et al. (2013) detected 40,000 MgII absorbers 
in the SDSS DR7 data set with a completeness of $> 95\%$. However, it usually takes a long time (weeks to months)
 to detect the numerous MgII absorbers in the large SDSS quasar spectral data. In order to detect more MgII absorbers in the growing SDSS quasar
 spectral data from the latest SDSS data releases for statistical study of MgII absorbers and their properties, 
a computationally fast algorithm with high detection efficiency would 
be extremely useful. 

The traditional method of detecting the MgII doublet usually includes three steps. First, quasar spectra are
 fitted and normalized. It is common to fit the continuum of the quasar spectra with a spline function and 
broad emission lines with Gaussian profile functions. 
 However, this process sometimes fails to give a good continuum fit, such as erroneously fitting the 
 spiky emission lines. Also it requires lots of human interference. Principle Component Analysis (PCA), 
as a sub-branch of machine learning, can also perform the continuum fitting. However, it takes many hours to
 solve the covariance matrix and one may need to carefully choose how many eigen vectors to use to
 fit the quasar continuum. There is a trade-off between algorithm computational speed and result accuracy and the goodness of the continuum fitting has a large effect on the following line detection in a large dataset. After the spectra are normalized, a sliding multi-line modeling is 
performed on the normalized quasar spectra to search for absorption lines. 
If the potential lines can be identified and reach a certain signal-to-noise ratio (SNR),
 they are considered as robust absorption lines. One major disadvantage of the traditional method is
 that every quasar spectrum in the data set requires the continuum fitting and multi-line modeling, 
resulting in long computation time. It usually takes weeks to months to finish the entire data set of a SDSS data release
 with many  human interferences. 

Due to the several aforementioned drawbacks of the traditional method, we employ a deep neural network based method
 that can alleviate these issues, including significantly reducing detection time and improving detection accuracy and its adaptiveness to other narrow absorption line detection scenarios.
In this neural network, the computation units
 consist of a set of neurons. These neurons are connected by three basic components: an input layer, hidden layers,
 and an output layer. Many efforts have been made in the deep neural network community
 to improve the computational efficiency in the hidden layers. 
Currently, there are two popular neural networks capable of data classification. One is the 
convolutional neural network, which focuses on classification
 suitable for Mg II absorber identification in quasar spectra.
 The other neural network is
 called the recurrent neural network, which focuses on time series data analysis. In the domain of machine learning,
 the deep neural network has proven to be an effective method to solve all kinds of big data problems in industrial fields.
For instance, Krizhevsky et al. (2012) trained a deep neural network and
 classified 1.3 million images with very high accuracy. This was the first time that the deep neural network
 outperformed the previous classification algorithms 
in both accuracy and computation time. He et al. (2015) presented a residual learning framework with 152 layers
 and won 1st place for the ImageNet Large Scale Visual Recognition Challenge 2015 classification task. The success of the residual network indicates that complicated problems can be solved by very deep neural networks.

The deep neural network based methods exhibit the following merits that benefit astronomy researchers: 1) Large data 
sets empower accurate representation learning. Deep neural network based methods perform best with rich training data. The curve fitting of  quasar continuum and absorption line profiles in narrow absorption line detection has actually repeatedly generated the training samples. By carefully designing the neural network, we are able to train an accurate deep neural network model for classification. 2) Easy to use and share. With a trained model, each prediction on newly added observations is just a feed forward
 process. Trained models could also be easily shared among researchers.
 
In astronomy, deep neural networks have been 
proven to be powerful in three astronomy fields: 1-D spectra, photometry or light curve classification, 2-D image
 classification, and parameters tuning in numerical simulation.
In the 1-D astronomical data field, the first application
 of a deep neural network was spectral classification in a SDSS data set (Hala et al. 2014). This study showed that
the convolutional neural network is capable of classifying  quasar spectra, stellar spectra and galaxy spectra with 
an accuracy of 95\%. Graff et al. (2016) improved the Swift trigger algorithm for long Gamma Ray Bursts (GRBs) 
with the random forest and neural network. 
Yuan et al. (2016) first developed a deep neural network to identify the broad dust absorption line systems in 
SDSS quasar spectra. This neural network contains 3 convolutional layers and one pooling layer. The network reached 99\% accuracy in identifying strong 2175 \AA\  broad absorption bumps in simulated quasar spectra. 
Recently, Park et al. (2017) applied a neural network in detecting damped Lyman-$\alpha$ broad absorption
 lines with high accuracy. Hampton et al (2017) employed a neural network to decrease the amount of time required for human interaction in emission line studies. Deep neural networks have also shown great potential in other 1-D data analysis such as pulsar 
signal detection (Bethapudi et al. 2017), asteroseismology classification  (Hon et al. 2017), earth-like exoplanet
 detection (Pearson et al. 2017),  transit source detection (Wright et al. 2017), and redshift estimation 
(Isanto et al. 2017).

 In the 2-D astronomical data field, deep neural networks are very useful in star-galaxy 
classification and galaxy morphology study. The deep neural network method was first introduced in this 2-D field in 1996. 
Bertin et al. (1996) used a neural network in star-galaxy classification and implemented this method in Source-Extractor (Sextractor, Bertin et al. 1996).  
Kim et al. (2016) used a deep neural network to do star-galaxy classification and reached an accuracy of  99\%.
 Petrillo et al.(2017) applied a deep neural network in gravitational lens detection. The deep neural network method can
 also help significantly reduce false detections in large datasets. For instance,  
Kuntzer et al. (2017) used a deep neural network  to 
do  stellar classification with diffraction patterns on simulated images with a very high success rate. 
 In the numerical simulation field, Mustafa et al. (2017) used a generative adversarial network to infer model
 parameters. The deep neural network method has also been used in cosmological simulations and
 radio surveys to constrain parameters (e.g., Sullivan et al. 2017).

In this paper, we use a deep neural network to identify narrow absorption line systems of MgII $\lambda \lambda$
 2796, 2803\AA\ doublets in the SDSS DR12 data set (Zhu et al. 2013). The description of the dataset and preprocess is presented 
in Section 2. We discuss the structure of our neural network and hyper parameter tuning in Section 3.
 We present our results in Section 4 and draw conclusions and discuss future improvement of algorithms in Section 5.

\section{Data set and Preprocess}

It is important to generate artificial datasets based on the physics behind the data 
 for training the neural network since most published dataset cannot reach an accuracy of 100\%. However, it is very difficult to create an artificial dataset
 with all the parameters matching with the real dataset. In this work, we use both artificial spectra based 
on the DR7 quasar sample and  the DR7 MgII catalog as the training sets, and the DR12 MgII catalog 
as the test set to investigate  how the neural network responds to Mg II absorption lines at different redshifts and equivalent widths.

\subsection{Real MgII Catalog As The Training Set}

In this paper, we use four catalogs, all of them can be downloaded  at
 \url{http://www.guangtunbenzhu.com/jhu-sdss-metal-absorber-catalog}. 1) The MgII absorber DR7 catalog.
 This catalog is generated by the pipeline described in Zhu et al. (2013). It contains 47734 MgII absorbers at
 $0.35< z_{abs} <2.28$ among a total of 26761 quasar spectra.
 We only use the strong MgII absorbers with $EW_{2796/2803} \geqslant 3 \sigma_{EW2796/2803}$ as real detections.
 This leaves 22985 quasar spectra with strong Mg II absorbers. We label this catalog as catalog1. 
2) The quasar catalog in DR7. This catalog is originally from the SDSS quasar catalog (Schneider et al. 2010) and
 contains 84533 quasar spectra. We label it as catalog2. 3) The MgII absorber DR12 catalog, which has 24064 quasar spectra
 with strong MgII absorbers using the above criteria. We label it as catalog3. 4) The quasar catalog in DR12, which
 contains 57479 quasar spectra. We label it as catalog4.

Before we pre-process individual spectra in both the DR7 and DR12 catalogs, we cross-matched catalog2 and catalog4 with published quasar catalogs (Schneider et al. 2010, Paris et al. 2017). For the DR7 data set, we found that  83962 out of 84533 quasar spectra have been found in catalog2 and 22919 out 22985 quasar spectra have been found in catalog1. For the DR12 data set, we found that  51174 out of 57479 quasar spectra have been found in catalog4 and 21267 out 22985 quasar spectra have been found in catalog3

We use the DR7 datasets (catalog1 and catalog2) as the training set and use the DR12 datasets (catalog3 and catalog4)
 as the test set. In both datasets, we label strong MgII absorption spectra as 1 and 
label the spectra with weak Mg II absorption lines or no MgII absorption lines as 0.
 In the DR7 dataset, the positive sample contains 22919 spectra and
the  negative sample contains 61043 spectra. We randomly select 22919 spectra
 out of 61043 no-MgII spectra and label them as 0 as the control sample. Previous studies show that the most
 optimal way to train a deep neural network is to have equal size for both positive and negative samples 
(Hensman 2015). We  use the bootstrap method here to derive the sample error from the neural network by repeating
 the random selection step 5 times. This results in 5 training sets
 which contains the same MgII spectra but different control spectra. Then we randomly divide each training 
set into the training dataset and validation dataset with the ratio of 4:1.

The quasar spectra in both datasets are preprocessed. First, bad pixels in the spectra have been masked out.
 Since we only focus on the relative strength of the absorption lines, each spectrum has been rescaled by 
using the small continuum region without strong emission line or broad absorption line contamination. For instance, 
we use the region around $4150$\AA\ $\sim 4250$\AA\ in quasar spectra to rescale the spectra
 with quasar emission redshift, $Z_{emi} \leqslant 1.0$. 
We adopt  the regions around 3020\AA\ $\sim 3100$\AA, 2150\AA\ $\sim 2250$\AA\ and 1420\AA\ $\sim 1500$\AA\
 in the quasar's rest frame for quasar spectra with redshift $1.0 < Z_{emi} \leqslant 1.8$, $1.8 < Z_{emi} \leqslant 2.8$ 
and  $2.8 < Z_{emi} \leqslant 4.8$, respectively. After the spectrum has been rescaled, 
we use interpolation to rebin the spectrum to produce the final spectrum. All the spectra should have the same dimension of 3841 pixels.  Lastly, we only search for MgII absorbers between the redward of the CIV emission line and the blueward
 of MgII emission line. We assign 0 for the pixels outside this range.  One example is shown in Figure 1.

\subsection{Artificial MgII Spectra As The Training Set}

We used the entire quasar catalog in Schneider et al. (2010) to build our artificial dataset. 
This catalog includes 105783 quasar spectra. Since the search window is between 1500\AA\ and 2900\AA\
 in the quasar's rest frame, we only use the quasar spectra with the redshift range from 0.3 to 2.8.
 This left a sample of 80000 spectra. In order to simulate quasar spectra with artificial 
MgII absorbers, we first conduct continuum fitting to each spectrum, inject an artificial
 Mg II absorber to each continuum as the second step, then add noises to create a final simulated spectrum.

The Principle Component Analysis (PCA) method is used to fit the continuum of each quasar spectrum.
 To save computation time, we used the scikit-learn python package (Pedregosa et al 2011) to perform the PCA fitting in an iterative way (IPCA; Budavari et al. 2009). Specifically, we first divide the entire quasar sample into subsamples by
 their emission redshifts with a bin size of 0.2. In each subsample, the total number of spectra
 is less than 10000. We then performed the IPCA to each subsample and derived their 
corresponding eigen-spectra. We found that each continuum can be well fitted using a combination 
of 20 eigen spectra. 

After spectrum continuum fitting, we inserted MgII absorption profiles into the
 continuum spectra. We first identified spectral regions
 where S/N is greater than 3.0  using the original SDSS quasar spectrum and error array.
 We then randomly inserted the MgII doublet absorption lines into these regions 
with a line significance level (SL) greater than 3 $\sigma$, calculated from the local S/N, where $\sigma$ 
is the equivalent width error for the simulated absorption line. The simulated absorption profile
 is drawn from the full width at half maximum (FWHM) distribution of observed MgII
 absorption lines. In this study, we randomly selected 5000 MgII absorbers from the DR7 MgII
 catalog and measured their FWHMs. The measured FWHMs follow a Gaussian distribution with
 a mean of $\mu = 1.7$\AA\ and $\sigma = 0.7$\AA. We used this distribution to generate 
FWHMs for artificial MgII absorbers. After we inserted the MgII profiles into the quasar
 continuum, we added noise into the spectra using the error array in the original 
quasar spectrum. Figure 2 illustrates the entire procedure. We generated 80000 simulated
 quasar spectra with MgII absorption and randomly selected 50000 spectra with SL greater than 3 $\sigma$. Since 
 some of the weak absorbers we inserted (e.g. $EW_{2796/2803} < 0.3 $\AA ) will be
 suppressed by the noise, we also filtered out the weak absorbers with $EW_{2796/2803} < 0.3 $\AA. 
 Compared to the equivalent width distribution of the real DR7 dataset, we found the equivalent width distribution of
 our simulated spectra generally follow the trend of real DR7 dataset except that there are less strong
 absorbers ($EW_{2796/2803} > 2.5 $\AA\ ) in the simulated data set, as shown in Figure 3. But this is unlikely to lead to a significant difference in the training results since the strong absorption is very easy to identify.  As shown in  the bottom panel of Figure 7, almost all the strong absorbers can be successfully detected. We treated these simulated spectra as the positive sample in the training dataset. We also randomly generated 50000 spectra without any absorption lines and treated them as the negative sample. Both the positive sample and negative sample were passed through the same processes before being used for training and testing our neural network.

\section{Structure of the Network and Hyperparameter Tuning}

The deep neural network used in this study has two essential algorithms:
 the feed-forward calculation and error back propagation. There are three basic components in the neural network:
 an input layer, hidden layers, and an output layer. The input layer is the feature vector 
 $\bm{S}=(s_{0}, s_{1}, s_{2}, s_{3}...)$. Since a quasar spectrum has a dimension of 3841, 
the input feature vector has the same dimension of the quasar spectrum. Each element in the
 input vector is the flux measured in each wavelength bin. Each value from the input layer 
is duplicated and sent to all of the hidden nodes in the hidden layers.

Convolutional layers are key elements in the Convolutional neural networks (CNNs). For the structure 
of hidden layers used in this work, we employ convolutional layers and pooling layers to boost classification accuracy
 of our neural network. In the fully connected network, the nodes in hidden layers of 
the convolutional network are called filters. In each node, the length-fixed filter slides across
 the input matrix with a fixed set of weights. In other words, the input image is convolved with 
the filter and applied into a non-linear activation function. The output matrix from each node 
is called a feature map. If we denote the feature map from the k-th node as $A^{k}$, whose filters are
 determined by the weights $W^{k}$ and bias $b_{k}$. The feature map $A^{k}$ can be obtained by:
\begin{equation}
A^{k}_{ij}=\sigma((W^{k} * s)_{ij}+b_{k}).
\end{equation}
Here $\sigma$ is a non-linear transformation that functions as the ``activation'' that transforms the input feature vector
 into an output vector. The rectified linear unit $\sigma(x)=max(0,x)$ (ReLU; Nair \& Hinton 2010) 
and the hyperbolic tangent unit  $\sigma(x)=tanh(x)$ are popular activation functions used 
in the hidden layers. Here we used the ReLU function as the activation function in the hidden layers.

Each feature map can capture a local spatial feature of the data in the previous layer. In the traditional
 convolution neural network, the network not only contains the convolutional layers but also contains
 the pooling layers. The basic usage of pooling layers is to reduce the spatial size from the previous
 layer while preserving the most useful information. It is also referred to the downsampling layer. 
There are many different pooling techniques, like maxpooling, average pooling and L2 pooling. 
In this work, we only apply the maxpooling technique. The basic idea of a maxpooling layer is that
 it applies a filter with a certain size and stride to the input image, and outputs the maximum number
 in every subregion that the filter convolves around. Another type of layer we use in this work 
is the dropout layer. The idea of dropout layer is that this layer will drop out a random set of activations 
in that layer by setting them to zero. This will prevent overfitting during the training.

In this work, we perform a binary classification since we want to identify spectra with 
MgII absorption lines from other quasar spectra. For a binary classification case,  
 we use the sigmoid function in the output layer to describe the possibility of 
an output vector $\bm{a} = \sigma(x)$:
\begin{equation}
\sigma(x)=\frac{1}{1+e^{-x}},
\end{equation}
where $\bm{x}$ is the output from previous layers.

In order to improve the performance of the neural network, we aim to find an algorithm 
which can optimize the weights $\bm{W}$ in every layer so that the output vector $\bm{a}$ 
from the network approximates the ground truth $\bm{y}$ for all training inputs $\bm{x}$. 
Since the neural network aims to perform the binary classification, the likelihood function
 of binary classification is the Bernoulli distribution:
\begin{equation}
L(\bm{W})=a^{y}(1-a)^{1-y}.
\end{equation}
Instead of maximizing the log likelihood function, we minimize the cost function 
$C(\bm{W}) = - log L(\bm{W})$:
\begin{equation}
C(\bm{W})=-y \log a-(1-y) \log(1-a).
\end{equation}
This is the cost function to be optimized. The gradient descent algorithm is employed to minimize the equaiton 4.
 The gradient descent updates the weights by using:
\begin{equation}
w_{i} \rightarrow w^{'}_{i}= w_{i} -\eta \frac{\partial L}{\partial w_{i}},
\end{equation}
where $\eta$ is called the learning rate. The gradients are computed via the back-propagation algorithm 
(Rumelhart et al. 1986). Since the stochastic gradient descent method is a relatively fast optimization 
algorithm, we first use this method to optimize the cost function for
 different neural network configurations and examine which model is better by comparing their test accuracies. After we determine 
the model configuration, we use the adaptive moment estimation \textit{Adam} (Kingma et al. 2015) 
 algorithm to achieve a better result.

Our CNN is implemented in Python 2.7 using the open-source libraries Keras and TensorFlow. The training 
of the CNN is executed on 1 GeForce Tesla M40 GPU. A deep learning classifier usually has multiple stacks
 of neural network layers on top of one another as its structure. In this paper, we also built a neural network with multiple hidden layers. The muti-layer structure contains a vast
 combination of free parameters (hyper-parameters). It is very difficult to optimize the neural 
network since it has a large number of free parameters to determine, like numbers of layers, 
types of layers, filter size, strides etc. Bengio et al. (2012) introduced some empirical methods 
about network parameter tuning, but this greatly depends on the dataset. Another method of hyperparameters tuning is called grid search,
 but it is impractical if a training set is very large as this method is very computationally expensive. 
The typical size of a narrow absorption line feature is around 50 pixels in the spectra and
 using a smaller filter size (3 or 5) in the first layer make it harder and slower for the neural network to identify narrow absorption features from noise compared to using larger filters.

In this study, we trained our neural network on both an artificial spectral dataset and
 a real spectral dataset. The configuration of the best model from our parameter fine tuning
 contains 10 layers, as shown in Figure 4. It has 5 convolutional layers, 3 maxpooling
 layers, and 2 fully connected layers. The input spectrum has 3841 pixels, while the output
 is transformed to a 96 feature spectrum, each of which has 766 pixels. Afterwards the input 
spectrum is convolved with 96 kernels which have a filter size of 15. {We found that the filter size of the first layer plays a very important role in the hyperparameter fine tuning. A first layer with a small filter window (less than 5 pixels) or with a large filter window (greater than 17 pixels) led to an accuracy below $70\%$. After being convolved 
with other convolutional and maxpooling layers, the specrum is shrunk to have 25 pixels.
 Then it is connected to two fully connected layers. The cost function we aim to optimize
 is equation (4). The training curves are shown in Figure 5. 
 The training accuracies are a little bit higher than the cross validation accuracies.
 This is largely due to the overfitting of CNN. Compared to training curves on the real dataset,
 the training curves of the artificial dataset took more time to converge.  There are two reasons for this difference. First, considering the artificial dataset is twice as large as the real spectrum dataset, it would take the neural network longer time to find a 
global minimum. Second, the fraction of strong absorption lines (EW > 1 \AA) is larger than the one in artificial set, so it took more time for neural network to achieve a good score on artificial set. Training on the real dataset reaches an accuracy of $93 \%$ on average, while training on the artificial dataset reaches an accuracy of $96 \%$.

In order to demonstrate that the neural network performs better when the first layer filter has a similar
 size as the absorption feature instead of having complex structures, we also adopt three ``popular"  models among the industrial domain: the VGG net, Inception net and Residual net. These models have good performance when dealing with real world problems. The common feature of these neural networks is that the window size in their first layer is relatively small.
 
 First we test the VGG net. The striking feature of this model is that it is much deeper than the typical network 
but has a very small filter size (a typical size is 2 or 3). The accuracy is 82\%, as shown in Table 1. 
The second model we test is the Inception net. In the Inception model, each layer has a multi-filter size with a typical
 size between 2 and 5 which allow them to capture the feature on different scales. 
The Inception net has an accuracy of 78.7\%. The final model we compare is the residual net.
 This network has small filter sizes, but deeper layers. 
The accuracy of the residual net is only 66.8\%. These networks all have small filter
 sizes at the first layer (typically 2 or 3). After being trained on real dataset for 20 epochs,
 the final accuracy never exceeds an accuracy of $\sim 85\%$. All of these performances are worse than that achieved by our custom designed network with the first layer filter size closely matching Mg II absorption lines. In order to further investigate the correlation between the filter size parameter and the detection accuracy, we increased the first layer filter size in our model, VGG model and Inception model and track the accuracy on the validation set. As shown in Figure 6, we can find that the accuracy for each model gradually reach the maximum when the filter size close to the feature size.

\section{Results}
\subsection{Evaluation Metrics}
In this paper we adopt standard classfier performance metrics to evaluate our deep neural  
network performance. The metrics used to describe classifier performance are defined as follows:

\noindent
\textbf{Accuracy}: The number of correct predictions out of all predictions.

\noindent
\textbf{Precision(P)}: The ratio of correct predictions of positive sample to all made predictions towards the positive sample.

\noindent
\textbf{Recall(R)}: The ratio of correct predictions of positive sample to all targets truly in the positive sample.

\noindent
\textbf{F1 Score}: The harmonic mean of precision and recall with 1 as a perfect score.

\noindent
\textbf{ROC AUC}: Receiver Operating Characteristic's Area Under Curve, which measures the neural network's
 average performance across all possible score thresholds. It has a value of 1 for a perfect classifier.
\subsection{Results on DR12 dataset}
The DR12 dataset contains 21267 strong MgII absorber spectra out of 51174 quasar spectra. Each spectrum has
 been preprocessed by the same method introduced in section 2.1. The test set contains 21267 positive samples 
(spectra with strong MgII absorption) and  29907 negative samples (spectra without MgII absorption). 
Since we have five training sets for both artificial datasets and real datasets with the same 
positive sample and a different negative sample selected from DR7 quasar data, we individually trained the neural network with five training sets. After the training step, we average the 
detection accuracies on the DR12 quasar data for all the training datasets. It took the computer 
9.3 seconds and 9.7 seconds on average over the five training sets to analyze the 51174 spectra and identify all of the strong Mg II absorbers using 
the neural networks trained with DR7 quasar spectra and artificial data, respectively.
Assuming that the positive and negative labels in the DR12 dataset generated from Zhu et al. (2013) 
pipeline are 100\% accurate, our overall performance is very good on the DR12 
dataset using the neural networks based on real and artificial data training. 
Table 2 lists performance metrics. The average accuracy is $\sim 90\%$ and  $\sim 86\%$
achieved with the real data trained network and the artificial data trained network, respectively.
The small accuracy difference between the real data trained model and artificial data trained model is likely caused by the accuracy of the published MgII catalog used in generating the real training set; in other words, the labels in the real training data are not $100\%$ accurate.

To analyze our new results, we first investigate statistical  properties of the MgII sample identified
 by the deep neural networks. Figure 7 shows the successfully detected spectra for both models in the 
$Z_{emi}$, $Z_{abs}$ and equivalent widths spaces. The overall distributions in different parameter
 spaces are consistent with those drawn from Zhu et al (2013)'s sample. This means that the MgII lines 
detected by the deep neural networks can draw the same scientific statistical results as those drawn 
from the previous study using the traditional method to detect MgII absorbers. Nevertheless, 
we were able to find small differences between the previously detected Mg II absorber sample and our
 detected sample. Most of the missing spectra are located at low 
equivalent widths for both models. Our current neural network models have difficulty in  
detecting MgII absorbers in very noisy spectral regions or weak MgII absorption features.
 In addition, our current models perform relatively poor at the high $Z_{emi}$ regions. 
This is likely due to poor sky subtraction of quasar spectra at long wavelengths where Mg II absorbers are searched 
because the searching window between 1550\AA\ and 2800\AA\ in the quasar rest frame 
is shifted towards long wavelengths for high redshift quasars. 
 This issue is also noticed in the $Z_{abs}$ distribution. MgII absorbers at high $Z_{abs}$
 tend to be harder to detect than the ones at low $Z_{abs}$.

Two of the successfully classified examples (a true positive detection and a true negative detection)
 based on the artificial spectra trained model are shown in Figure 8. 
 This model can easily detect spectra with strong MgII absorption lines and
 most of the true negatives. We further diagnose the falsely detected sample.
 Two of the representative examples are also shown in Figure 8. There are some challenges that 
may affect metrics performance. First, some of the MgII doublet features are not obvious
 in the spectra (as shown in the bottom left panel in Figure 8). The reason why the neural network
 fails to detect relatively weak MgII lines is that the neural network cannot distinguish which are the 
narrow weak absorption lines and which are noises unless the narrow absorption lines are much stronger
 than the local noises in the spectra. Second, the CNN model we developed cannot distinguish MgII 
absorption lines from other absorption lines. One example is shown at the bottom right panel in this figure.

We further investigated 
the true nature of the ``false positive'' and ``false negative'' samples identified by the artificial
 spectra trained model compared to that labeled otherwise in Zhu et al. (2013)'s catalog using our selection 
criteria of strong MgII absorbers. This time, we used the traditional pipeline to measure spectra in both samples. 
In the false positive sample of 4132 spectra, we were able to identify 2538 spectra with Mg II absorbers 
where signals can pass the $3\sigma$ detection, but are missed by Zhu et al (2013)'s pipeline.
 One of the examples missed by Zhu et al. (2013) is shown in Figure 9. 
In the false negative samples of 2803 spectra, 1382 spectra have true Mg II absorption lines
 which can pass the $3\sigma$ detection. This means the rest of MgII spectra in the false negative sample
 are either too weak or not real.  Based on the missed MgII sample and falsely detected MgII samples,
 we recalculated the accuracy of our neural network and the pipeline accuracy of Zhu et al. (2013). 
The true accuracy of our neural network is $94.1\%$ while the true accuracy of Zhu et al. (2013)
is  $92.3\%$. Our neural network 
has similar  accuracy to the traditional method adopted in Zhu et al. (2013)'s pipeline, 
but has reduced the computational time to find these absorbers
 by a few orders of magnitude over the traditional method. Here we built two catalogs for the DR12 sample. One is the MgII absorbers missed by Zhu et al. 2013 pipeline. The other one is the false detection by Zhu et al. (2013). All catalogs can be found here: \url{https://github.com/brainiac21/NAL_deepnets}.

\begin{table}
\centering

\label{my-label}
\begin{tabular}{|l|l|l|l|}
\hline
Model Name & VGG   & Inception & Resnet \\ \hline
Accuracy   & 0.822 & 0.787     & 0.668  \\ \hline
\end{tabular}
\caption{Accuracy of different popular deep neural network models trained with the real DR7
 MgII catalog over 30 epochs. All the popular  deep neural network
 models cannot reach a fairly high accuracy despite that all the models have much more parameters
 and more complex network structure than our model.}
\end{table}

\begin{table}
\centering
\label{my-label}
\scalebox{0.9}{%
\begin{tabular}{|l|l|l|l|l|l|}
\hline
Metric    & Real data traing set & Artificial data traing set  \\ \hline
Accuracy  & 0.8971     & 0.8649     \\ \hline
Precision & 0.8977    & 0.8165   \\ \hline
Recall    & 0.8524     & 0.8710        \\ \hline
F1 score  & 0.8729     & 0.8428      \\ \hline
ROC AUC   & 0.8907     & 0.8658      \\ \hline
\end{tabular}}
\caption{Metrics on the DR12 dataset for both deep neural network models trained with the artificial and real
 DR7 MgII spectra. Assuming that the strong Mg II absorbers identified by the traditional method by Zhu et al. (2013) 
are 100\% accurate, the trained neural network can reach $\sim 90\%$ detection accuracy on average
 with the real data trained model while reaching $\sim 86\%$ accuracy for the artificial data trained model. 
The computation time used in analyzing $\sim 50000$ spectra is $\sim 9$ seconds for both models.}
\end{table}


\begin{table}
\centering
\caption{Results from the deep neural network model trained with the artificial Mg II absorbers in DR7}
\label{my-label}
\begin{tabular}{ll}
True Positive & False Positive \\ \hline
17135         & 4132          \\ \hline
True Negative & False Negative \\ \hline
27104         & 2803  \\ \hline
\end{tabular}
\end{table}

\section{Conclusion and Discussion}

In this paper, we developed a convolutional neural network that can detect the strong 
MgII absorption lines ($EW_{2796/2803} > 0.3 $\AA) in SDSS DR12 quasar spectra
 with $\sim$94 \% accuracy. This neural network was trained on the artificial MgII sample using DR7 quasar spectra. 
This accuracy allows us to draw the same overall statistical results as that based on the strong MgII
 sample identified by the traditional method adopted in Zhu et al. 2013. However, the detection
 speed with our deep neural network method
 has been tremendously improved. It only takes 9 seconds with our neural network to completely 
analyze $\sim 50000$ quasar spectra and detect strong MgII absorbers in the quasar spectra, compared to weeks of time to 
detect these Mg II absorbers using the traditional method. This is the first time that the deep
 neural network has been applied to narrow absorption line detection. This neutral network has been 
applied to SDSS IV DR14 quasar spectra and was able to identify $\sim 50000$ new Mg II absorbers 
in 50 seconds. A paper on this new Mg II catalog will be published soon (Zhao et al. 2018, in preparation).

Unlike other popular neural network models which use small filter sizes for the first layer, 
our study shows that filter sizes in the first convolutional layer matching Mg II absorption feature sizes can produce the best detection accuracy. In our Mg II absorption spectra, two 
Mg II absorption lines typically occupy about 50 pixels. A filter size of $\sim 20$ pixels in the first layers can optimally capture the two absorption line characteristics that will be
 propagated through the rest of the deep neural network layers to eventually produce positive detections.
 On the other hand, it is difficult to distinguish an absorption line feature from random 
noises with a small filter size  while a very large filter size also loses its detection sensitivity.

We re-examined the DR12 MgII catalog selected by the JHU's pipeline (Zhu et al. 2013) and independently 
identified 2538 additional strong MgII absorbers. On the other hand, our neutral network misses a
 similar number (1648) of strong Mg II absorber in Zhu et al.'s catalog. This is largely due to either
 noisier spectral regions or spectral regions with numerous sky emission lines. The overall detection 
accuracies for both Zhu et al's traditional method and our neural network are similar, 92.3\% vs. 94.1\%. 
It is quite possible that our accuracy may be improved further by adopting the encoder technique 
commonly used in deep neural network applications to de-noise the quasar spectra. A followup paper 
on applying this de-noise method in both quasar spectra and Kepler data to improve the detection 
accuracy will be reported (Zhao et al. 2018b, in preparation).

Although our neural network can quickly identify a larger number of MgII absorbers in SDSS 
quasar spectra, this model has several areas for future improvement. For instance, it still needs
 the traditional method to derive absorber parameters such as equivalent width, redshifts, and FWHMs.
 Due to non-uniformity of Mg II absorber distribution in the parameter spaces,
 our detection accuracy may be improved by training
 the neural network on simulated data with Mg II absorbers drawn from the same parameter spaces.
 In our current training, we only adopted the same FWHM distribution as the observed one. 

During our training, we also noticed that the spectral feature, the Mg II absorption lines, 
only occupy about 50 pixels, while the neural network needs to learn the entire spectra with
 $\sim$3800 pixels, which is inefficient. Parks et al. (2017) used a sliding window to localize
 the Damped lyman-$\alpha$ lines and feed them into a neural network. This method can be adopted
 to our study. However, it would be time consuming since one spectrum would have to be divided 
into multiple pieces for the neural network to learn. We plan to explore this sliding window 
further in combination of the CNN design to look for an optimal way to quickly locate the absorbers'
 redshifts while minimizing its processing time.

\section{acknowledgement}
We thank the anonymous referee for comments and suggestions which have helped significantly improve the paper quality. 
We also thank Mr. Nolan Grieves for carefully proofreading the paper. 
Funding for SDSS-III has been provided by the Alfred P. Sloan Foundation, the Participating Institutions,
 the National Science Foundation, and the U.S. Department of Energy Office of Science. The SDSS-III
 web site is \url{http://www.sdss3.org/}.

SDSS-III is managed by the Astrophysical Research Consortium for the Participating Institutions of the
 SDSS-III Collaboration including the University of Arizona, the Brazilian Participation Group,
 Brookhaven National Laboratory, Carnegie Mellon University, University of Florida, the French 
Participation Group, the German Participation Group, Harvard University, the Instituto de 
Astrofisica de Canarias, the Michigan State/Notre Dame/JINA Participation Group, Johns Hopkins
 University, Lawrence Berkeley National Laboratory, Max Planck Institute for Astrophysics, 
Max Planck Institute for Extraterrestrial Physics, New Mexico State University, New York University,
 Ohio State University, Pennsylvania State University, University of Portsmouth, Princeton University,
 the Spanish Participation Group, University of Tokyo, University of Utah, Vanderbilt University,
 University of Virginia, University of Washington, and Yale University.

\begin{figure}
\centering
\includegraphics[angle=0,scale=.50]{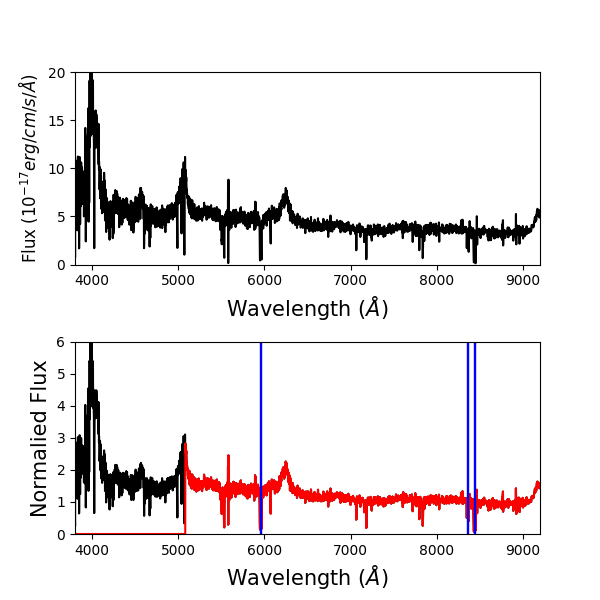}
\caption{Illustration of a preprocessed quasar spectrum. Top: The original MgII-quasar spectrum in the observer's frame. Bottom: The same quasar spectrum was rescaled and rebined in the observer's frame. The search window for the quasar spectrum is from the redside of CIV emission line to the blueside of MgII emission in quasar rest frame. The final spectrum used in neural network training is plotted in red. Three MgII absorption line systems are marked with blue lines.}
\end{figure}

\begin{figure*}
\centering
\includegraphics[angle=0,scale=.60]{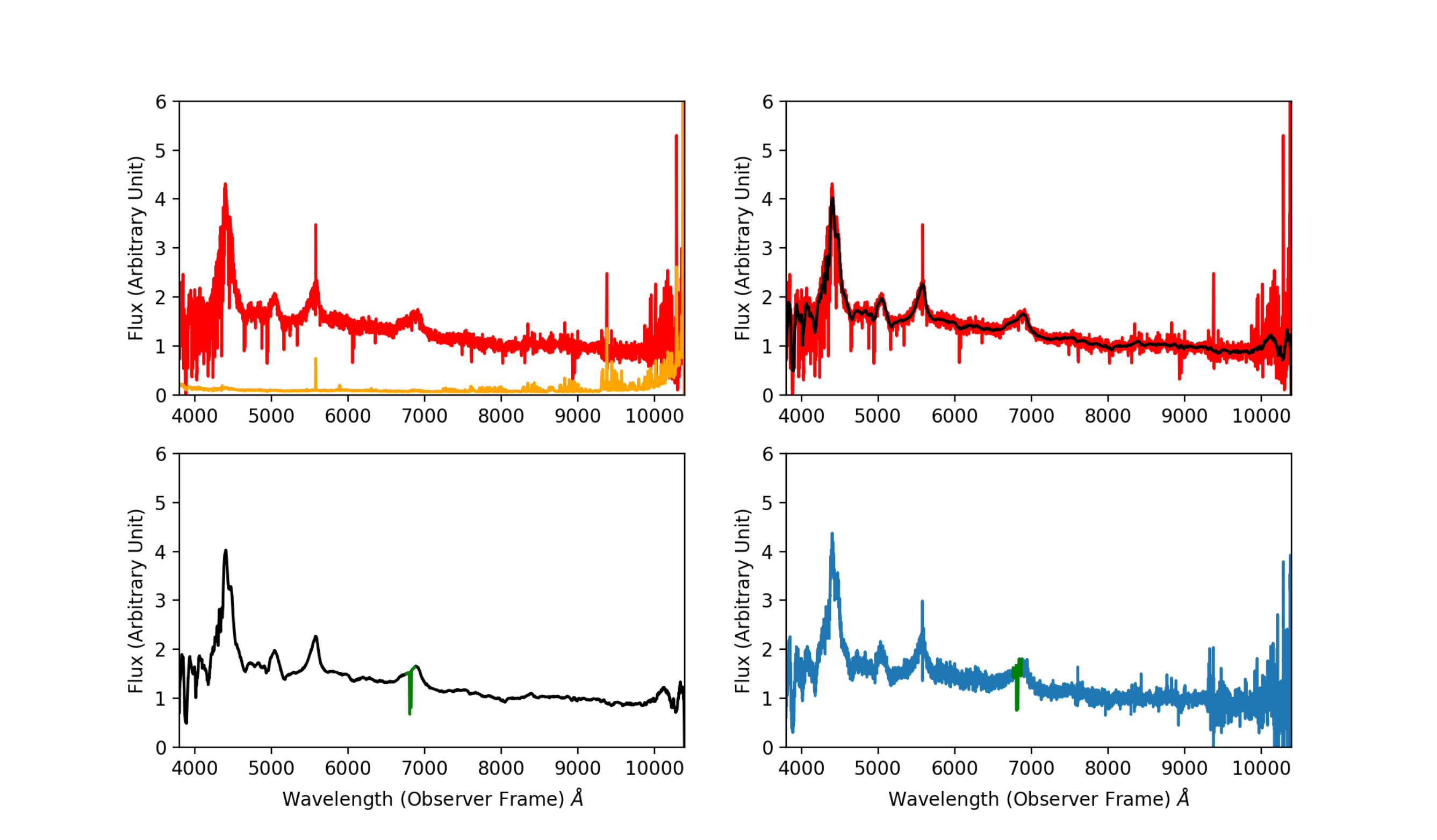}
\caption{Illustration of the artificial MgII generation procedure. Top Left: The original quasar spectrum. The error array is labeled with orange. Top Right:  Continuum fitting of the quasar spectrum with the PCA method. The continuum is shown in black. Bottom Left: The quasar continuum inserted with a MgII absorber. The inserted MgII absorber is labeled with green. Bottom Right: Gaussian noises from the original spectrum error array injected to the quasar continuum with a MgII absorber. The final spectrum in blue is used for for neural network training.}
\end{figure*}

\begin{figure*}
\centering
\includegraphics[angle=0,scale=.60]{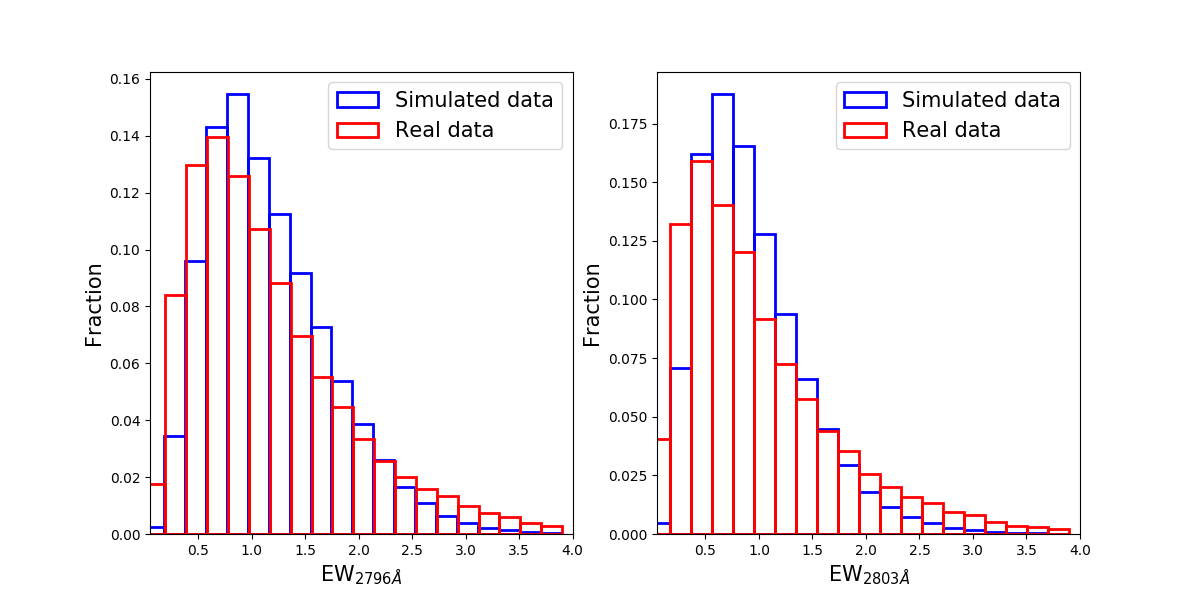}
\caption{Equivalent width distributions of the real dataset and simulated dataset. Left: Equivalent width distributions of $EW_{2796}$. The simulated data is labeled with blue and the real dataset is labeled with red. The overall trends for both datasets are the same except that the fraction of strong absorbers in the real dataset is larger than the simulated one. Right: Equivalent width distributions of $EW_{2803}$. The simulated data is labeled with blue and the real dataset is labeled with red.}
\end{figure*}

\begin{figure*}
\centering
\includegraphics[angle=0,scale=.40]{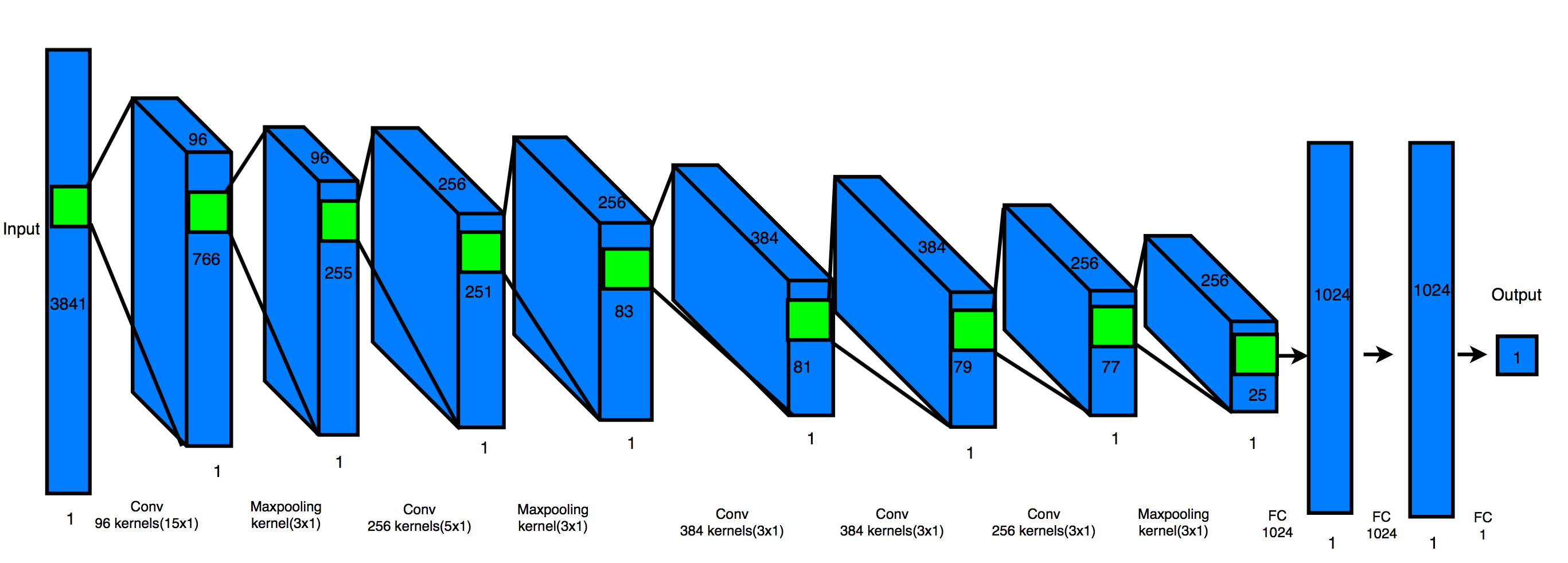}
\caption{The architecture of the convolutional neural network aiming at detecting the MgII narrow absorption lines. The input spectra have 3841 pixels.}
\end{figure*}

\begin{figure*}
\centering
\includegraphics[angle=0,scale=.60]{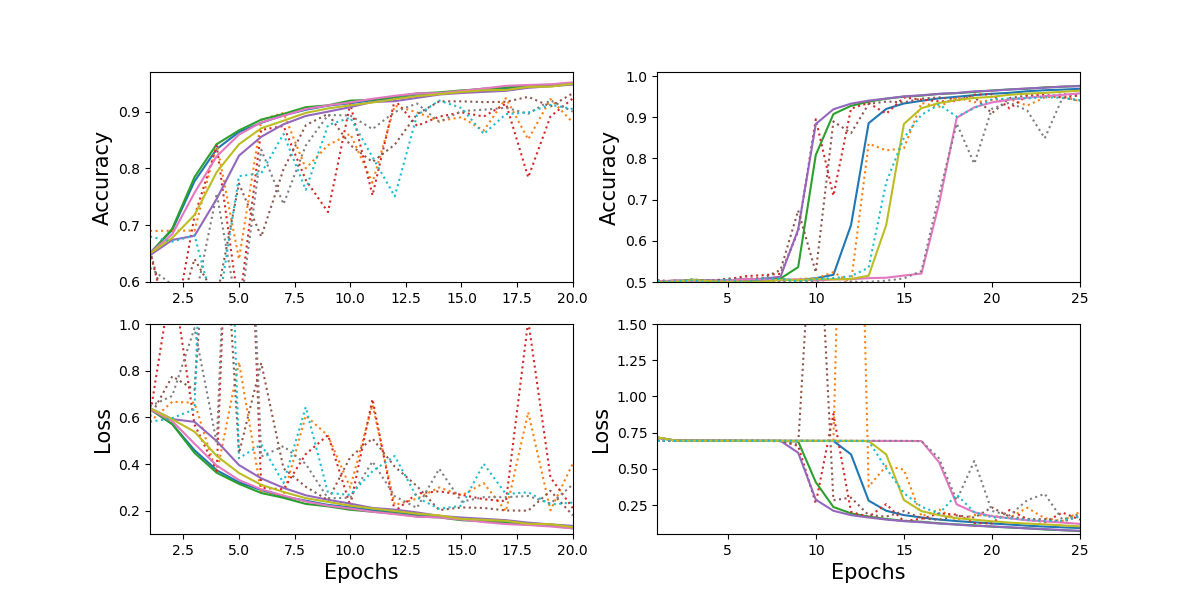}
\caption{Metrics as the function of training epochs. Top Left: training and validation accuracy as a function of training epochs for the real data trained model. Top Right: training and validation accuracy as a function of training epochs for the artificial data trained model. Bottom Left: training and validation loss as a function of training epochs for the real data trained model.  Bottom Right: training and validation loss as a function of training epochs for the artificial data trained model. The dotted lines in both panels are the metrics from the validation set and the solid lines are from the training set. We trained the convolutional neural network with the real data set over 20 epochs. Compared to the real data trained model, the artificial data trained model took longer time to reach the best performance. The final average of the validation accuracy for 5 training sets for the real data and artificial data are $\sim 93\%$ and $\sim 96\%$, respectively.}
\end{figure*}

\begin{figure*}
\centering
\includegraphics[angle=0,scale=.60]{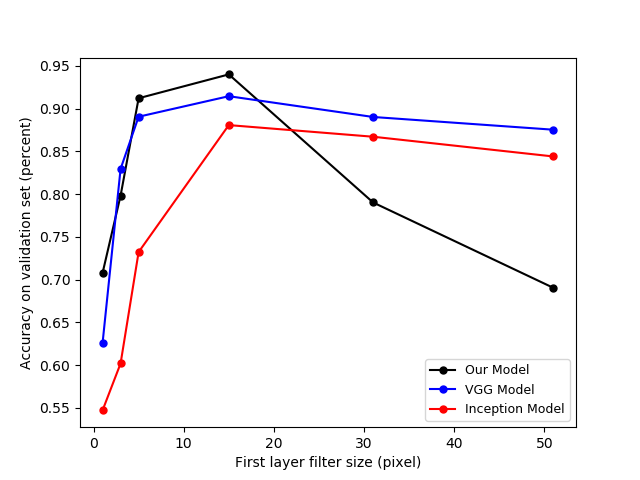}
\caption{The detection accuracy as the function of the filter size in the first layer in different neural network models. As the filter size increases, the accuracy of each model reach an maximum when the filter size is close to the absorption feature size. After the filter size is larger than the feature size, the accuracy of each model drops.}
\end{figure*}

\begin{figure*}
\centering
\includegraphics[angle=0,scale=.50]{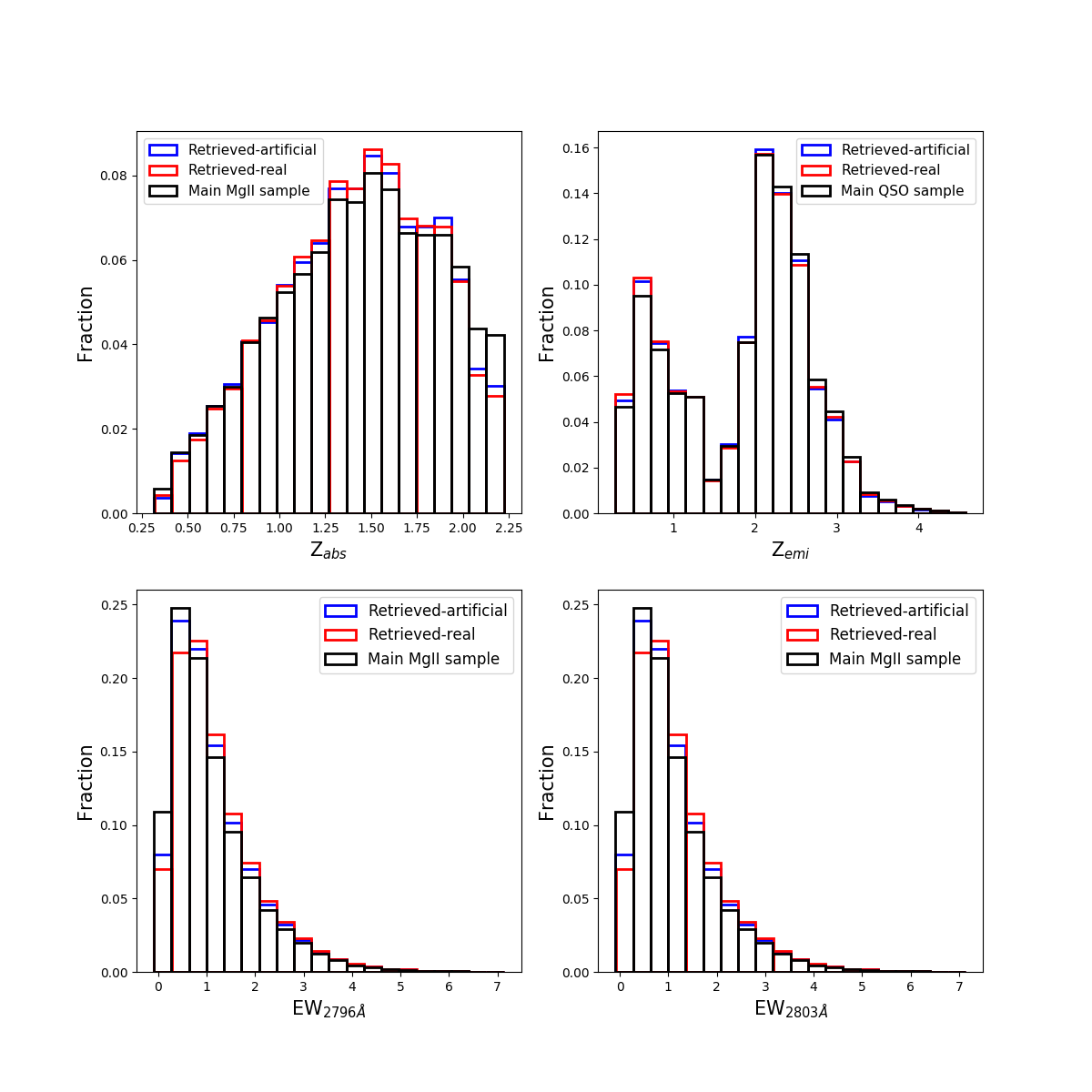}
\caption{Statistical results from the successfully classified sample in the DR12 dataset. In each panel, each distribution has been normalized. The results from the artificial data trained model and real data trained model are both shown here. Top Left:  $Z_{abs}$ distribution. By comparing with catalog3, most of the falsely classified samples appear at high redshift for both the artificial data trained and real data trained models. Top Right: $Z_{emi}$ distribution. By comparing with the main DR12 sample, most of the falsely classified samples also appear at high redshift for both the artificial data trained and real data trained models. Bottom Left: $EW_{2796\AA}$ distribution. The distribution of entire MgII sample is higher than the MgII sample detected by neural networks at smaller $EW_{2796 \AA}$ region. Bottom Right: $EW_{2803 \AA}$ distribution. The distribution of the entire MgII sample is higher than the detected MgII sample at smaller $EW_{2803 \AA}$ region.}
\end{figure*}

\begin{figure*}
\centering
\includegraphics[angle=0,scale=.40]{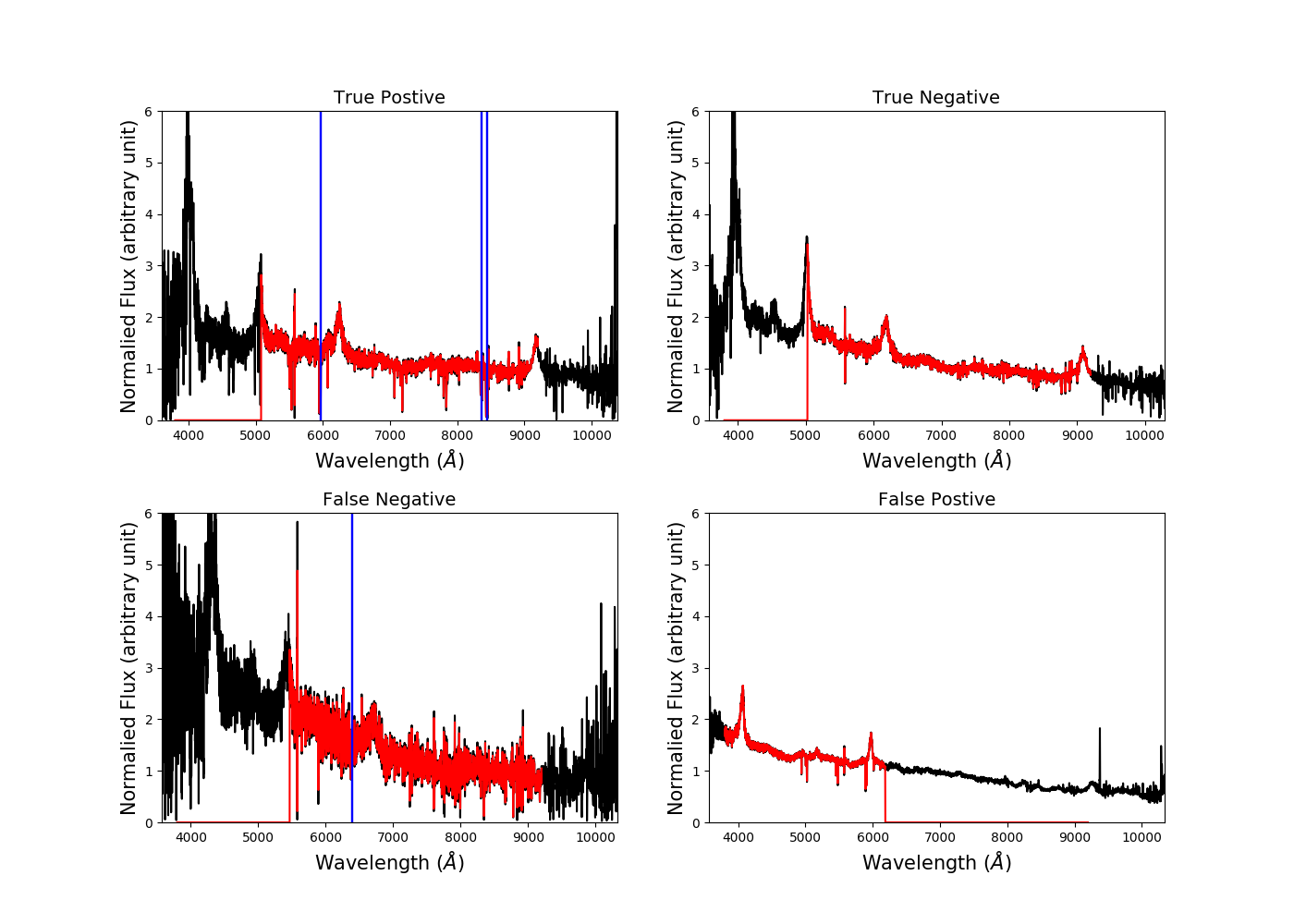}
\caption{Examples of quasar spectra in the observer's frame for the positive sample and negative sample. Top Left: true positive. This spectrum contains three MgII absorption line systems. The position of MgII absorption line systems are marked with blue lines. The search window is plotted in red. Top Right: true negative. This spectrum contains no narrow absorption lines. Bottom Left: false negative. This spectrum contains one MgII narrow absorption line system, but the signal to noise ratio of the line region is low making it hard for the CNN to detect them. Bottom Right: false positive example, This spectrum contains no MgII absorption line systems, but it contains other strong metal absorption lines which are mis-identifed by the neural network as MgII absorbers }
\end{figure*}

\begin{figure*}
\centering
\includegraphics[angle=0,scale=.60]{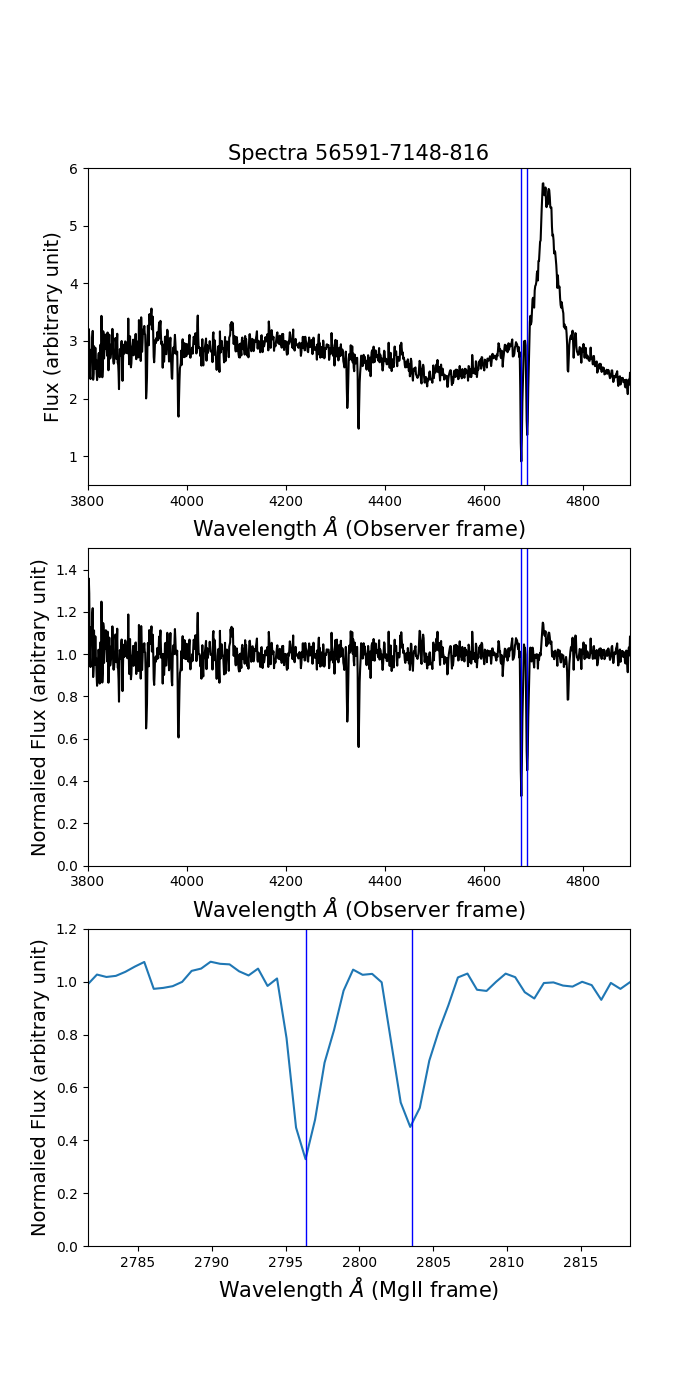}
\caption{One of the missed MgII absorbers by Zhu et al. (2013) pipeline. The spectrum is labeled with MJD-Plate-fiber ID. The top panel shows the original quasar spectrum. The missed MgII absorber is marked with two blue lines. The panel in the middle shows the normalized spectra with the MgII absorber marked by two blue lines. The bottom panel shows the zoom-in MgII absorber from the normalized spectra in the absorber's rest frame.}
\end{figure*}


\bsp	
\label{lastpage}
\end{document}